\documentclass[conference]{IEEEtran}

\usepackage{cite}

\usepackage[pdftex]{graphicx}
\graphicspath{MFigs/}

\DeclareGraphicsExtensions{.pdf,.jpeg,.png}
\usepackage{listings}
\usepackage{url}
\usepackage{listings}
\usepackage{pifont}
\let\labelindent\relax
\usepackage{enumitem}
\usepackage{subfigure}
\usepackage{color}
\usepackage{multirow}
\IEEEoverridecommandlockouts

\begin{document}
\title{PINPOINT: Efficient and Effective Resource Isolation for Mobile Security and Privacy\vspace{-2ex}}
\author{
\IEEEauthorblockN{Paul Ratazzi\IEEEauthorrefmark{1}\IEEEauthorrefmark{2}, Ashok Bommisetti\IEEEauthorrefmark{2}, Nian Ji\IEEEauthorrefmark{2} and Wenliang Du\IEEEauthorrefmark{2}\vspace{-2ex}}
\\
\IEEEauthorblockA{\IEEEauthorrefmark{1}Information Directorate, Air Force Research Laboratory, Rome, NY}
\IEEEauthorblockA{\IEEEauthorrefmark{2}Dept. of Electrical Engineering \& Computer Science, Syracuse University, Syracuse, NY}

\thanks{Approved for public release; distribution unlimited (88ABW-2015-0958-20150316).}
}

\pagenumbering{arabic}

\maketitle

\thispagestyle{plain}
\pagestyle{plain}

\begin{abstract}
\boldmath
Virtualization is frequently used to isolate untrusted processes and control their access to sensitive resources.  However, isolation usually carries a price in terms of less resource sharing and reduced inter-process communication. In an open architecture such as Android, this price and its impact on performance, usability, and transparency must be carefully considered. Although previous efforts in developing general-purpose isolation solutions have shown that some of these negative side-effects can be mitigated, doing so involves overcoming significant design challenges by incorporating numerous additional platform complexities not directly related to improved security. Thus, the general purpose solutions become inefficient and burdensome if the end-user has only specific security goals.

In this paper, we present \textit{PINPOINT}, a resource isolation strategy that forgoes general-purpose solutions in favor of a ``building block'' approach that addresses specific end-user security goals. PINPOINT embodies the concept of Linux Namespace lightweight isolation, but does so in the Android Framework by guiding the security designer towards isolation points that are contextually close to the resource(s) that need to be isolated. This strategy allows the rest of the Framework to function fully as intended, transparently.  We demonstrate our strategy with a case study on Android System Services, and show four applications of PINPOINTed system services functioning with unmodified market apps. Our evaluation results show that practical security and privacy advantages can be gained using our approach, without inducing the problematic side-effects that other general-purpose designs must address.
\end{abstract}


\section{Introduction}
\label{sec:intro}

Over the decade since its introduction, Android has been a stunning success, eclipsing the market share of every other mobile operating system by a huge margin, and now shipping on well over \textit{1 billion} new devices annually\cite{Gartner:2692318:2014}.  This growth, however, has not been without its pains. By one recent measure, 97\% of today's mobile malware targets the Android operating system \cite{FSecureThreat2013}. Commensurate with this trend, end-users are increasingly concerned about privacy and protecting personal information. Unsurprisingly though, the typical end-user possesses little or none of the specialized technical expertise necessary to fully understand the security implications of installing apps, granting permissions, entering sensitive data, etc. As a result, most users have a hard time using currently-available security indicators to identify which apps they should trust, and which they should not \cite{Chin:2012:MUC:2335356.2335358}. Faced with this lack of knowledge and confusing choices, many become complacent or careless in performing what amounts to critical system administration tasks.

Even though successive releases of Android continue to enhance and improve security \cite{SecEnhance}, balancing security with usability has proven difficult. For example, Android 4.3's App Ops feature for selective permission granting was hidden from end-users in version 4.4.2, apparently due to usability concerns. Because developers cannot anticipate the endless security configurations App Ops makes possible, many apps failed to function or simply crashed when their permissions were selectively revoked by the user\cite{EFFAppOps13}. Although this functionality is still obtainable using third-party apps, most uninformed users are unlikely to make the additional effort of activating and understanding these features at the level of technical detail necessary. As a result, even the latest releases of Android 5.0 do little to help end-users protect themselves from software they choose to install.

While many users may not fully understand the technical aspects of security architectures, permissions, access control mechanisms, or measuring trust, most have no trouble articulating which high level objects, resources or capabilities they are most concerned with. For example, it's common to find users worried about how some apps might misuse location, sensitive data such as personal contacts, or personally-identifiable information (PII) like phone number or International Mobile Station Equipment Identity (IMEI). In response to this, numerous solutions to address these concerns have been proposed, and many of these involve some form of virtualization combined with access control to isolate untrusted applications.

Although every approach to isolation has its own unique strengths and weaknesses, all include trade-offs in terms of sharing and communication. In Android's open architecture, where resource sharing and inter-process communication (IPC) are fundamental to the platform's basic operation and usability, careful attention must be paid to fully understanding how a particular isolation boundary impacts the system's functionality and performance. If this trade-off is not considered at the outset of a design, significant performance, usability, and functionality issues can arise. Countering these negative side-effects requires designers to overcome challenging system problems, typically resulting in substantial modifications to the operating system, and significant second-order complexities not directly related to the initial security goals. These problems are especially prevalent in general-purpose designs that attempt to provide isolation containers for entire apps or virtual phones, without the benefit of \textit{a priori} knowledge of specific threat(s) or end-user security goals.

In this paper, we present \textit{PINPOINT}, a resource isolation strategy that forgoes general-purpose solutions in favor of a ``building block'' approach that addresses specific end-user security goals. By addressing stated security goals and no more, PINPOINT yields an effective result using only the minimum amount of isolation. This helps minimize or eliminate the negative side-effects that are sure to emerge when large parts of Android's open architecture are subject to isolation. The PINPOINT concept and its scope of applicability in Android are introduced in Section \ref{sec:concept}. Section \ref{sec:design} describes a case study whereby we implemented the PINPOINT concept as a lightweight \textit{hypovisor}\footnote{We use the term \textit{hypo}visor to indicate the relatively small scope of authority compared with \textit{super}visors (i.e., kernels) that have authority over an entire userspace, and type I (native) \textit{hyper}visors that have authority over one or more guest operating systems. The term has been used similarly in \cite{krishnan2014android}.} within Android's Context Manager in order to make possible isolation of any system service. Sections \ref{sec:application} and \ref{sec:evaluation} contain the implementation details and evaluation results of isolating four Android system services using this hypovisor. We summarize related work in Section \ref{sec:related_work}, future directions in Section \ref{sec:future}, and conclude the paper in Section \ref{sec:conclusion}.

\section{Motivations \& Concept}
\label{sec:concept}

\begin{figure}[!t]
\centering
\includegraphics[width=3in]{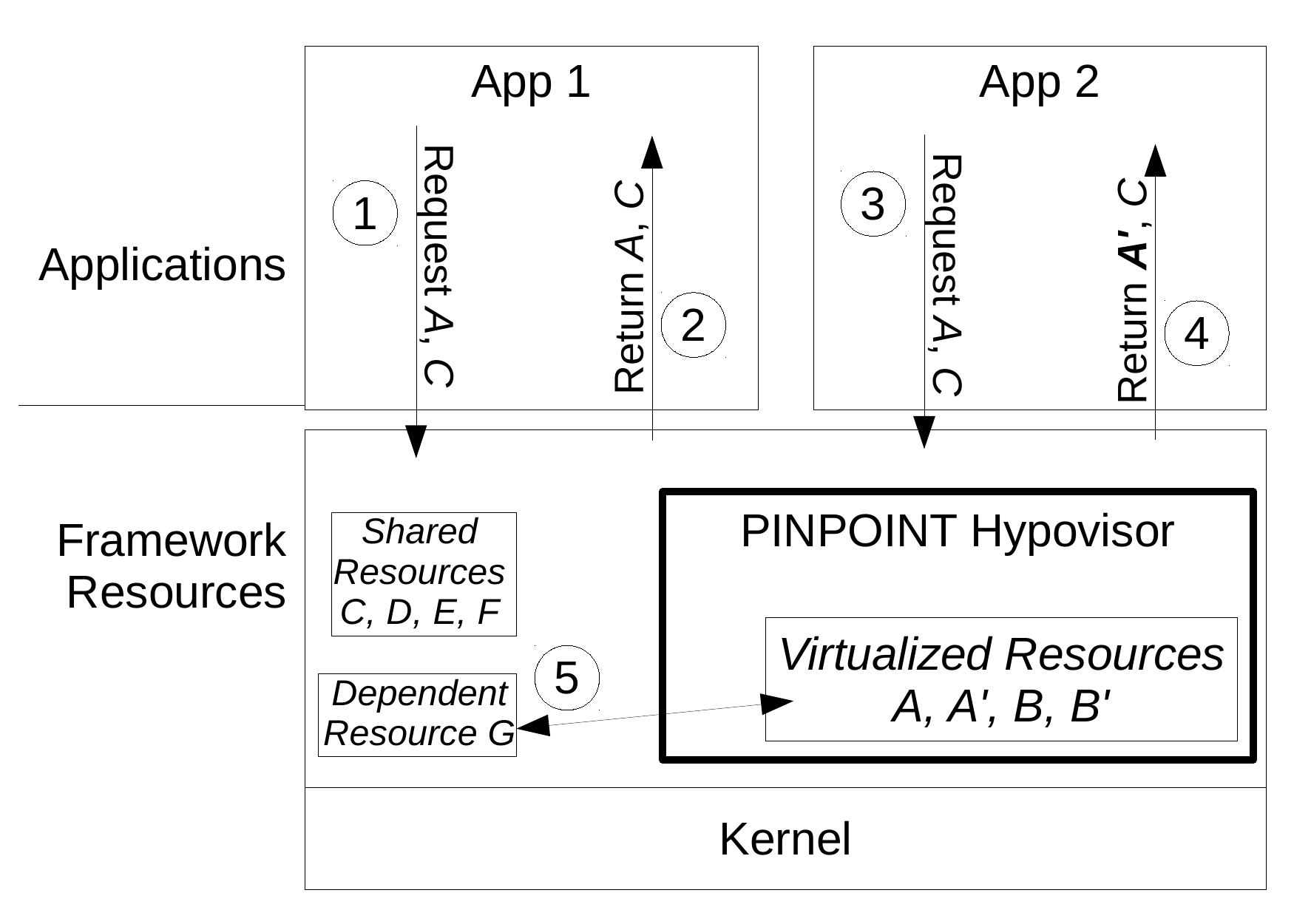}
\caption{PINPOINT concept showing minimized isolation to address security goals, with maximized sharing of system objects.}
\label{fig:concept}
\end{figure}

Our idea for PINPOINT originated from the realization that much of the time and effort devoted to implementing current isolation architectures (summarized in Section \ref{sec:related_work}) is spent on overcoming the negative side-effects on the Android system introduced by the chosen point of isolation. Most times, eliminating or mitigating these side effects requires challenging and far-reaching operating system modifications. Even when these challenges are met, we found ourselves unsure whether end-users would tolerate the remaining reductions in functionality, convenience, and performance. Furthermore, we thought it to be unlikely that these complex and less-usable designs would be adopted by Google or original equipment manufacturers (OEM), since Android user experience and its open architecture are paramount.

Nonetheless, we were intrigued by these designs' use of Linux Namespace lightweight isolation and were inspired to further explore the possibilities. Our conclusion is that for certain security scenarios, lightweight isolation using Linux Namespaces has many interesting advantages over heavyweight forms of isolation, such as virtual machines. Through a systematic analysis of Linux Namespaces, we identified six key traits that have value to our goals of providing effective yet efficient security. These traits are summarized in Table \ref{table:traits}.

\begin{table*}[!t]
\renewcommand{\arraystretch}{1.3}
\caption{Summary of Namespace Traits and Their Value to Android Security}
\label{table:traits}
\centering
\begin{tabular}{c||c}
\hline
\bfseries Namespace Trait & \bfseries Value to Android Security\\
\hline\hline
Fine-grained isolation of specific resources & Tailored isolation environment for each application, addressing specific threat and/or user goal\\
\hline
Resource-centric isolation & Match user perspective on security; increase usability; simplicity\\
\hline
High efficiency & Negligible performance impact; design simplicity\\
\hline
Share-by-default & Preserve open system design; avoid breaking things unrelated to the isolated resource\\
\hline
Transparent to host and apps & System retains control over apps; apps run unmodified\\
\hline
Small footprint (files, memory) & Little impact on performance \& resources; OTA updates\\
\end{tabular}
\end{table*}

On the other hand, we believe that the authors' direct use of Linux Namespaces as the point of isolation underlying the Android Framework breaks several basic assumptions of Android's open design. Hence, this choice of kernel-level isolation is the root cause of many of the complex system problems and negative side-effects they encountered. Although choosing an existing low-level mechanism as the foundation for isolation enabled their general-purpose isolation containers, there are many times when the cost of such comprehensive isolation is not worth it, especially when the end-user's security goals are relatively simple. For example, preventing a single untrusted app from accessing the device's IMEI. Our work fills this gap by providing an approach that \textbf{moves the point of isolation as close as possible to the object(s) needing isolation}, based on the stated security goals. In essence, we strive to realize the benefits of lightweight isolation in Android by identifying strategic locations where Linux Namespace \textit{concepts} can be implemented. As we will show, the result is a simpler implementation free of problems and far-flung platform side-effects. When specific security goals are taken into account, the result can be just as effective as general-purpose solutions.

Figure \ref{fig:concept} depicts our general, high-level concept. Given a specific security goal that relates to App 2's interaction with objects $A$ and $B$, our goal is to place the point of isolation (i.e., the hypovisor) at a strategic location that enables virtualization of only these objects such that App 1 and App 2 see different instances of each, while everything else about the system is common and unmodified. When trusted App 1 requests $A$ and $C$ (\ding{172}), the hypovisor returns instances of $A$ and $C$ (\ding{173}). On the other hand, when untrusted App 2 presents the same request (\ding{174}), the hypovisor returns instances of $A'$ and $C$ (\ding{175}). Since $C$ and other resources $D$, $E$, $F$ and $G$ are not related to the security goal, they are not virtualized, and either app may share them. Thus, the isolation size is minimized to just $A$ according to the threat, and resources that can and must be shared for transparent operation remain as shared as intended. Moreover, Framework complexities that would arise from utilizing kernel-level isolation mechanisms are completely avoided.

A non-trivial challenge that can sometimes arise when PINPOINTing certain resources is when there are other operating system components or resources, shown in Figure \ref{fig:concept} as resource $G$, that depend on interactions (\ding{176}) with $A$ or $B$ and are unaware that now multiple virtual copies of them exist. In these cases, $G$ must also be modified to account for this. Since this represents additional complexity, one must always consider whether this extra complexity will negate the lightweight benefits of the PINPOINT approach. If $A$ is a large and complex object that has many dependencies throughout the system, it's likely that creating virtual copies of $A$ will break many things that assume there is only one $A$. In cases like these, it may be better to use a coarser isolation such as those discussed in Section \ref{sec:related_work}. On the other hand, if $A$ has few dependencies, then the modifications to $G$ (if in fact there are any) will be straightforward. Two of our four case study applications described in Section \ref{sec:application} exhibit this characteristic, and these details will be included there.

Another challenge that can arise is when the same sensitive information can be revealed by more than one object. For example, let's say both $A$ and $C$ are capable of returning a piece of sensitive data such as IMEI. It is important that all of these paths be identified, and either blocked or added to the isolation boundary. Our case study encountered one example of this which will be described in Section \ref{sec:application}.

When designing and implementing the hypovisor, care must be taken to ensure that the system does not allow any form of delegation of the hypovisor's duties. For example, if the hypovisor is responsible for dispatching a capability, there must be no other ways for an entity to acquire that capability. Any other ways must be blocked in order to maintain the integrity of the isolation. Section \ref{subsec:security} contains a specific example of this and how we addressed it using mandatory access controls (MAC).

Identifying the best place to instantiate the hypovisor is key to achieving a balance between flexibility and specificity. In our experiences thus far, we have found that the best isolation points are places in the Framework where classes of objects and/or their capabilities are managed or dispatched to apps. In Android, many resources are abstracted as system services, and their capabilities are dispatched by \textit{ContextManager} (i.e., the native \texttt{servicemanager} process). As such, our initial work has focused on PINPOINTing system services, and our accomplishments thus far in this regard are described in Sections \ref{sec:design} and \ref{sec:application}. However, we see future opportunities for implementing complementary hypovisors in key places other than system services, including:

\begin{enumerate}
  \item High level data objects, such as \textit{ContentProvider}. These may leak personal data \cite{DBLP:journals/corr/LiuDZLZ14}.
  \item Binder and Intents. These may be used as a path for a malicious app to attack or trick other apps\cite{Felt:2011:PRA:2028067.2028089}.
  \item Camera, audio. These have obvious privacy implications if miused, e.g., \cite{Simon:2013:PSI:2516760.2516770}.
  \item Clipboard. Can be used as an attack channel \cite{ZhangClipboard}.
  \item Accessibility subsystem. May be malicious toward critical apps \cite{Jang:2014:AAE:2660267.2660295}.
  \item Notifications. Potential misuse \cite{179504}.
\end{enumerate}

\section{Case Study on Android System Service}
\label{sec:design}

In order to evaluate the PINPOINT concept, we undertook a case study using Android's system services framework in both Android 4.4.4 (KitKat) and 5.1 (Lollipop) on a Nexus 5 device. Since a wide variety of key resources are abstracted as system services, this choice illustrates that if the point of isolation is wisely chosen, a single PINPOINT hypovisor can be used for a variety of situations. To illustrate this point, we first provide some background on system services.

\subsection{Android System Services}
\label{subsec:system_services}

Interactions between Android applications and system services is enabled by the \textit{Binder} and \textit{Service Manager} subsystems. Binder relies on capability-based security and implements a ``call by invitation'' mechanism to allow communication among apps, system services and Service Manager.  As such, before an app is allowed to call a service, it must receive an invitation in the form of an \texttt{IBinder} token.

Invitations are first created when services are registered with the central directory of services known as \textit{Context Manager}. By design, there can be only one \textit{Context Manager}, a designation granted exclusively to the native \texttt{servicemanager} process early during the boot process, by way of its privileged relationship with Binder.\footnote{In fact, in the Android source (\path{frameworks/native/cmds/servicemanager/binder.h}), \textit{Service Manager} is described as ``The One Magic Object''.}  Once \textit{Service Manager} becomes \textit{Context Manage}r, \textit{System Server} registers core system services using the \texttt{addService()} method of \textit{Service Manager}. The result of this registration process is that \textit{Service Manager} now holds an invitation (\texttt{IBinder}) for every system service running on the device. When an app needs an invitation for one of these services, it contacts \textit{Context Manager}. \textit{Context Manager} then passes this invitation to the app, and upon seeing this transaction, \textit{Binder} updates its protected list of invitations held by the app. Invitations cannot be forged because any forged invitation will not have a corresponding entry in the protected list maintained by \textit{Binder}.

All requests for system services, even those made by system components, must go through \textit{Context Manager}. Thanks to Binder, the native \texttt{servicemanager} process has access to the trusted identity of the caller, in the form of the Linux \texttt{uid}, which corresponds to the Android \texttt{userId} and \texttt{appId}. This makes \texttt{servicemanager} an excellent place to implement a system service hypovisor that can regulate applications' interactions with virtualized system services. In this way, this hypovisor represents the PINPOINT ``sweet spot'' of being specific enough to limit inter-dependencies with other parts of the system, but flexible enough to apply to a large class of objects and the mechanism whereby their capabilities are dispatched.

\subsection{PINPOINTing System Services}
\label{subsec:arch}

\begin{figure}[!t]
  \centering
  \includegraphics[height=3in]{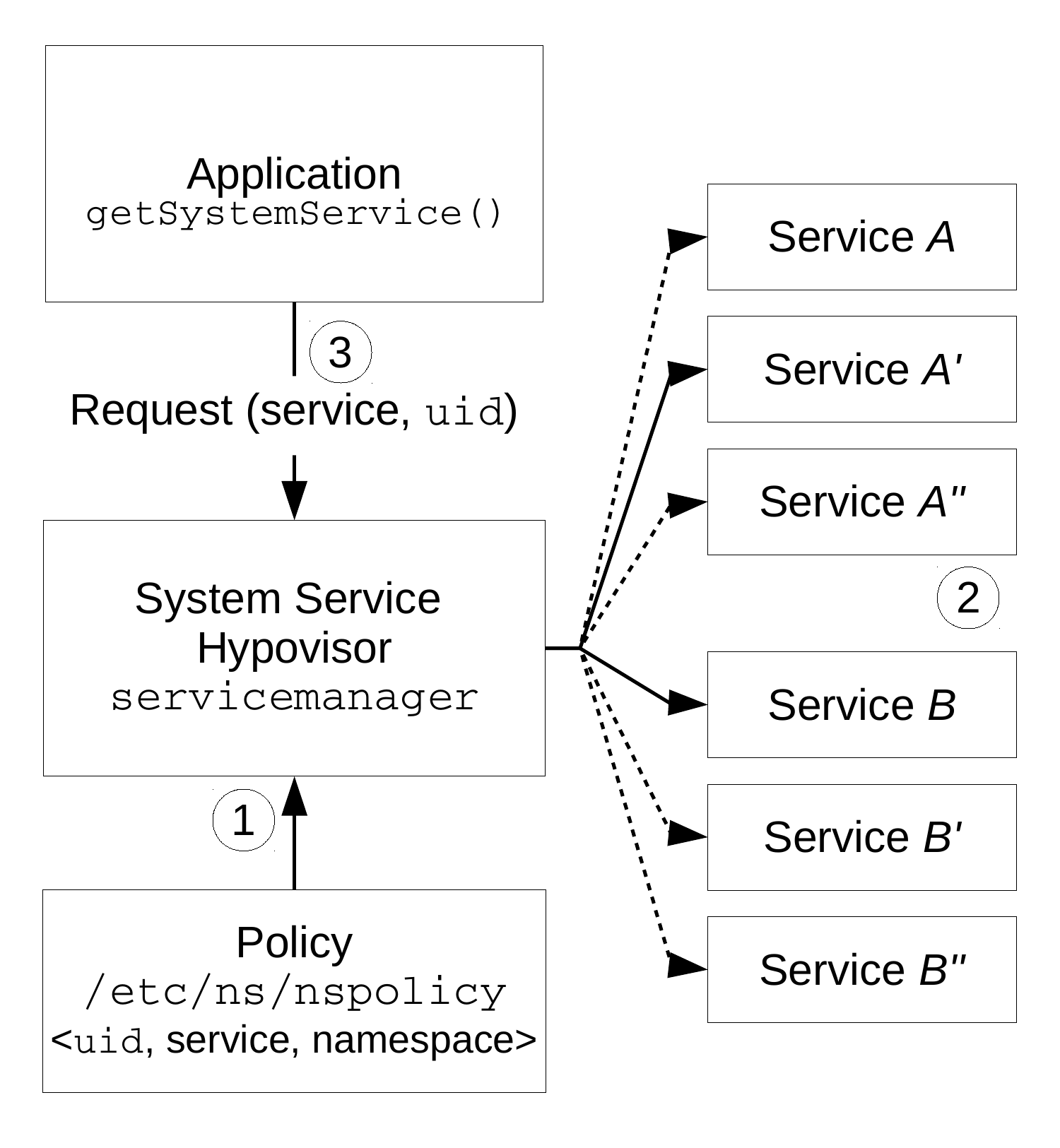}
  \caption{Design overview showing the service hypovisor and policy definition \ding{172}, virtual service plug-ins \ding{173}, and application \ding{174}.}
  \label{fig:design}
\end{figure}

We present our design in three parts as shown in Figure \ref{fig:design}, beginning with the central enabling core, the system service hypovisor, labeled \ding{172}. The hypovisor exists within the native \texttt{servicemanager} process, where all service lookups are processed and capabilities dispatched. Lookup requests by apps (\ding{174}) arrive in the form of a \textit{Binder} transaction containing the name of the service requested (e.g., \texttt{location}) and are identified by the app's Linux \texttt{uid} and \texttt{pid}. This identification can be trusted because it is applied by the kernel in the \textit{Binder} driver. By consulting with a secure policy file, \texttt{servicemanager} uses the \texttt{uid} to determine if the app has been assigned to an isolated namespace for the requested service. This policy is defined by a set of 3-tuples of $<uid, service\_name, namespace>$, where $uid$ corresponds to the \texttt{appId} of the app assigned to $service\_name$ namespace $namespace$. If the \texttt{uid} appears in the policy, then \texttt{servicemanager} replies with the handle of the virtualized service instance corresponding to the assigned namespace. If not, then its reply contains the handle of the global instance, as it would normally. The policy may specify more than one $service\_name$ and $namespace$ for a given $uid$ to contain apps that present a multi-dimensional threat. Since handle lookup requests can occur once or many times during the lifecyle of an app, the design also supports dynamic policy changes.

Currently, virtual services shown in Figure \ref{fig:design} at \ding{173} (e.g., $A'$, $A''$, $B'$ and $B''$) are preconfigured at build time and started by \textit{System Server} along with their global counterparts, $A$ and $B$. Typically, the virtual services have interfaces identical to the global service, but differ in terms of what they do. For example, the global location service returns the actual current location, while the other location services return noisy, random or preset locations via an identical public interface.

\subsection{Security Discussion}
\label{subsec:security}

In our PINPOINT case study, we introduce a lightweight services hypovisor into the native portion of \textit{Service Manager}. The purpose of the hypovisor is to isolate particular apps from various services as specified by the user's policy. This security discussion is included to provide a sense of the strength of this isolation. We begin with \textit{Binder}, since most of the isolation strength derives from \textit{Binder}'s security model.

Every process using \textit{Binder}, including system service threads within \texttt{system\_server} has a protected representation in the kernel as an instance of a \texttt{binder\_proc} structure. Each remote capability that a process holds is represented by one or more \texttt{binder\_node} structures attached to the \texttt{binder\_proc} instance. These nodes are known only to the kernel module and are used to determine the recipient of the communication, based on a handle provided from userspace. Handles are local references and mappings from handles to nodes are also stored securely in \texttt{binder\_proc}. Hence, only the kernel knows how to map a particular handle to the corresponding node.

When \textit{System Server} registers system services with \textit{Context Manager} using \texttt{addService()}, the kernel adds the service's \texttt{binder\_node} to \texttt{servicemanager}'s \texttt{binder\_proc}. \textit{Context Manager} also allocates a local index to each registered service. When an app asks \textit{Context Manager} for a handle to a service, \texttt{servicemanager} returns the handle and the kernel binder driver adds the service's \texttt{binder\_node} to the app's \texttt{binder\_proc}. Apps can also send handles they posses to other apps via Intent. Upon seeing the handle within the transaction, the kernel driver adds the node to the receiver's \texttt{binder\_proc} so that the recipient is now a valid holder of that capability. This is known as a \textit{binder transfer}.

With our addition of a services hypovisor, we do not change anything about \textit{how} handles are looked up and provided by \textit{Context Manager} or \textit{how} capabilities are propagated by way of the kernel binder driver. All apps, native and Java alike, are subject to the intervention of our hypovisor when requesting service handles. Any vulnerabilities in our prototype regarding \textit{how} service handles are obtained by apps, or vulnerabilities in the binder driver itself, are also vulnerabilities of stock Android and thus outside the scope of this discussion.

What our design does change is \textit{which} handles are given out. In PINPOINTing services, we have introduced the notion of remote service handles that should be unobtainable by certain apps. This is different than stock Android where \textit{Context Manager} acts as an open directory service, and obtaining a service handle via binder transfer from another app does not represent a capability leak. In our design, this rather unusual case of app-to-app transfers of system service handles must be prevented so that our hypovisor cannot be bypassed. Our prototype achieves this blocking in the kernel binder driver's \texttt{binder\_transaction()} function, through an extension of existing SEAndroid MAC. Specifically, we extend the \texttt{security\_binder\_transfer\_binder()} hook by also passing the \texttt{task\_struct} of the \texttt{binder\_ref} (for references) or \texttt{binder\_node} (for handles) being transferred, so that the hook function can extract the owner's security identifier (SID) and decide if the transfer should be allowed. Finally, we modified the type enforcement rules associated with \texttt{untrusted\_app} to disallow transfer of \texttt{u:r:system\_server:s0} binders between \texttt{u:r:untrusted\_app:s0} apps. By adding a \texttt{neverallow} rule, we further ensure at build-time that there are no \texttt{allow} rules elsewhere in the policy that are inconsistent with this. This effectively blocks any attempted bypass of our hypovisor, while allowing all other normal binder transfers among apps and the system to proceed.

\subsection{Policy Configuration}
\label{subsec:config}

As explained in Section \ref{subsec:arch}, the hypovisor within \texttt{servicemanager} consults a secure policy file to determine if the requester has been assigned to any virtual services. This policy can be created and updated by a variety of means: via the system Settings app, via launcher configuration, from hard-coded (i.e., build-time) mandatory policy, via over-the-air (OTA) updates in a mobile device management (MDM) architecture, etc. In our prototype, we included a default policy file in the system build, and updated it via \texttt{adb} and a custom launcher application. In terms of user-friendliness, our custom launcher enables the end-user to drag-and-drop app icons to and from different containers, each representing a specific PINPOINT configuration. For example, a particular container might be configured to protect two sensitive resources, location and IMEI, from the apps placed within it. When an app is dropped into this container, the launcher app automatically updates the policy with the \texttt{uid} and service names corresponding to the protected resources. This update takes effect immediately since \texttt{servicemanager} consults the policy each time the app makes a request.

\subsection{Limitations}
\label{subsec:limits}

Currently, our case study prototype requires all global and virtual system services to running whether or not any apps are assigned to them. In terms of overhead, this fact manifests itself as additional memory use by the \texttt{system\_server} process. Although we show in Section \ref{subsec:perf} that this overhead is small, we feel that this aspect of the design could be made more elegant and efficient in the future.

It is also important to note that our design does not provide full security domain isolation in the sense that it does not prevent apps from passing high-level sensitive information to other apps.
 
\section{Applications} 
\label{sec:application}

In this section, we describe the specifics of our experience with PINPOINTing four common system services, based on specific security goals. All implementations were tested in Android 4.4.4 (KitKat) and then ported to Android 5.1 (Lollipop) on a Nexus 5 device.

\begin{enumerate}
  \item \texttt{LocationManagerService}: A widely used location-finding service that binds with a number of abstract provider mechanisms. Security goal is to prevent untrusted apps from obtaining accurate location information\cite{de2013unique}. See Section \ref{subsec:location}.
  \item \texttt{IPhoneSubInfo}: A ``hidden'' service for accessing phone subscriber information, called only by other system services such as \texttt{TelephonyManager}. Security goal is to prevent untrusted apps from accessing sensitive subscriber information \cite{Enck:2011:SAA:2028067.2028088}.  See Section \ref{subsec:phone}.
  \item \texttt{InputMethodManagerService}: A service that arbitrates communications between apps and a variety of installed input methods, and has complex interactions with other system objects including \texttt{WindowManager}. Security goal is to protect critical apps from falling victim to malicious input methods\cite{6680023}. See Section \ref{subsec:ime}.
  \item \texttt{SensorService}: A native service that interfaces directly with hardware devices. Security goal is to prevent untrusted apps from obtaining accurate sensor data to steal data\cite{Xu:2012:TIU:2185448.2185465}\cite{Aviv:2012:PAS:2420950.2420957}, eavesdrop \cite{6680832}, or track movement/location\cite{Komeda:2014:UAR:2638728.2641299}. See Section \ref{subsec:sensor}.
\end{enumerate}

By and large, porting from 4.4.4 to 5.1 was straightforward. In one case however (\texttt{IPhoneSubInfo}), changes to the underlying service architecture required us to slightly redesign and expand the isolation boundary in order to continue to meet the security goal. This will be discussed below.

\subsection{Location Service} 
\label{subsec:location}

Although location services provide great convenience and enable new functionality for users, they have significant security and privacy implications if misused. While some apps require accurate location to fulfill their main purpose, others utilize location information only to enrich their primary function. For example, a social networking app's primary function is to interact with friends via photo and status updates. These apps usually enrich this interaction by attaching location to these updates. If the end-user wishes to prevent only this one app from knowing location, and still enjoy its primary friend-interaction functions, she must rely on the trustworthiness of the app's own settings and controls. This is because current location privacy support from Android itself is too coarse-grained to achieve the user's goal of isolating only this one aspect of this one app. If the app is poorly-written or malicious in its handling of location data, privacy leaks may occur despite the user's best efforts to prevent them. By PINPOINTing the location service resource, and placing only this app in the new location namespace, we can transparently and effectively address this user's security goal without inconveniencing her or introducing the complex system modifications and overhead of general-purpose solutions.

To demonstrate this, we PINPOINTed the location service to provide three separate location namespaces for assigning apps, each with different semantics but identical interfaces. The \textit{global location namespace} functions normally and is used with trusted apps. A \textit{fuzzy location namespace} provides reduced-accuracy location information by adding noise to location objects. Finally, a \textit{random location namespace} returns totally random location data to assigned apps.

\begin{figure}[!t]
\centering
\includegraphics[width=3in]{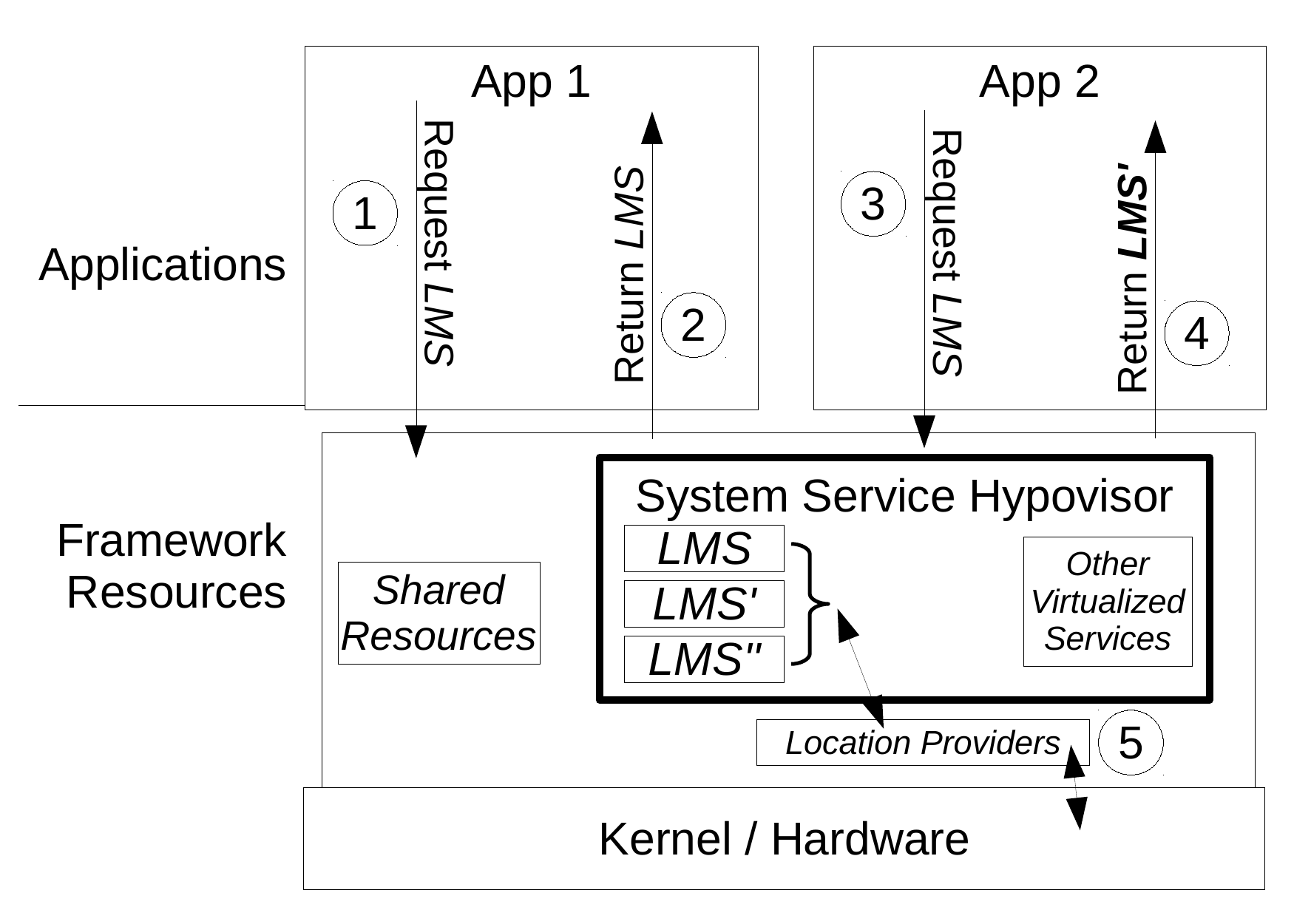}
\caption{PINPOINTing \texttt{LocationManagerService}.}
   \label{fig:locationServiceStructure}
\end{figure}

We implemented these two additional location namespaces by adding two additional system services, \texttt{LocationManagerService\_1} ($LMS'$) and \texttt{LocationManagerService\_2} ($LMS''$), as shown in Figure \ref{fig:locationServiceStructure}. These present the exact same API as the stock service, and thus are indistinguishable from the app's perspective.

Each location service binds to the standard set of common location providers such as \texttt{GpsLocationProvider} that interfaces through native code to actual hardware. However, as alluded to in Section \ref{sec:concept}, these providers represent dependent resources ($G$ in Figure \ref{fig:concept}) that are designed based on an assumption of only one location service. Thus, these must also be modified slightly to make callbacks to all three location services. Otherwise, $LMS'$ and $LMS''$ will never get location update callbacks since the providers are not otherwise aware of the virtualized services.

The semantics of the additional services are as follows: \texttt{LocationManagerService\_1} replaces location updates returned from the providers with random data, while \texttt{LocationManagerService\_2} adds random offsets to the same. Since each namespace is indistinguishable from the global location namespace in, apps in alternate namespaces behave normally and process the virtual location data as if it were real.

\begin{figure}[!t]
  \centering
  \subfigure[\textit{RunKeeper} running in location namespace with added noise.]{\label{fig:fuzzy_loc_ns}
    \includegraphics[height=2.5in]{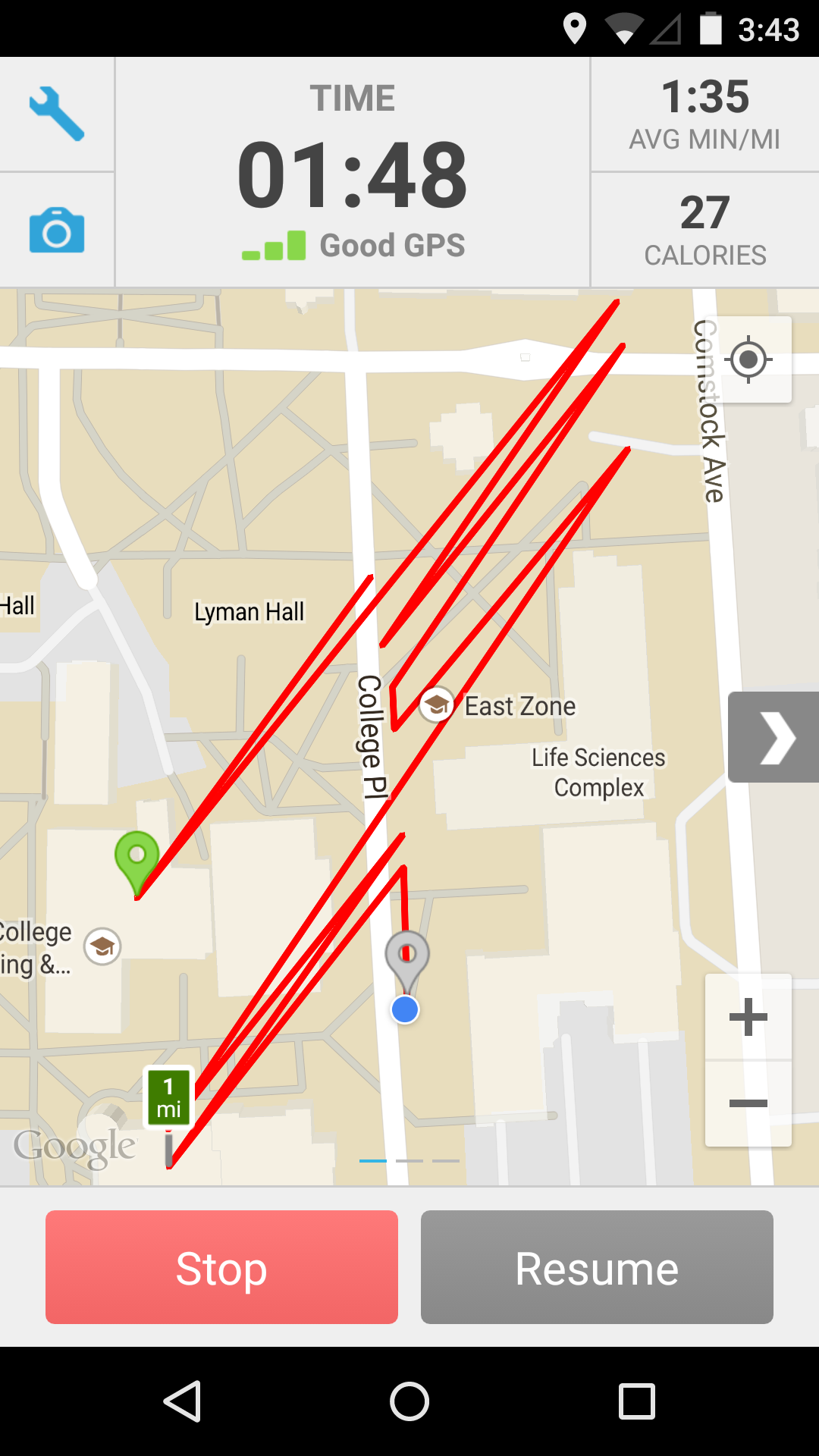}}
  \qquad
  \subfigure[\textit{RunKeeper} running in random location namespace.]{\label{fig:random_loc_ns}
    \includegraphics[height=2.5in]{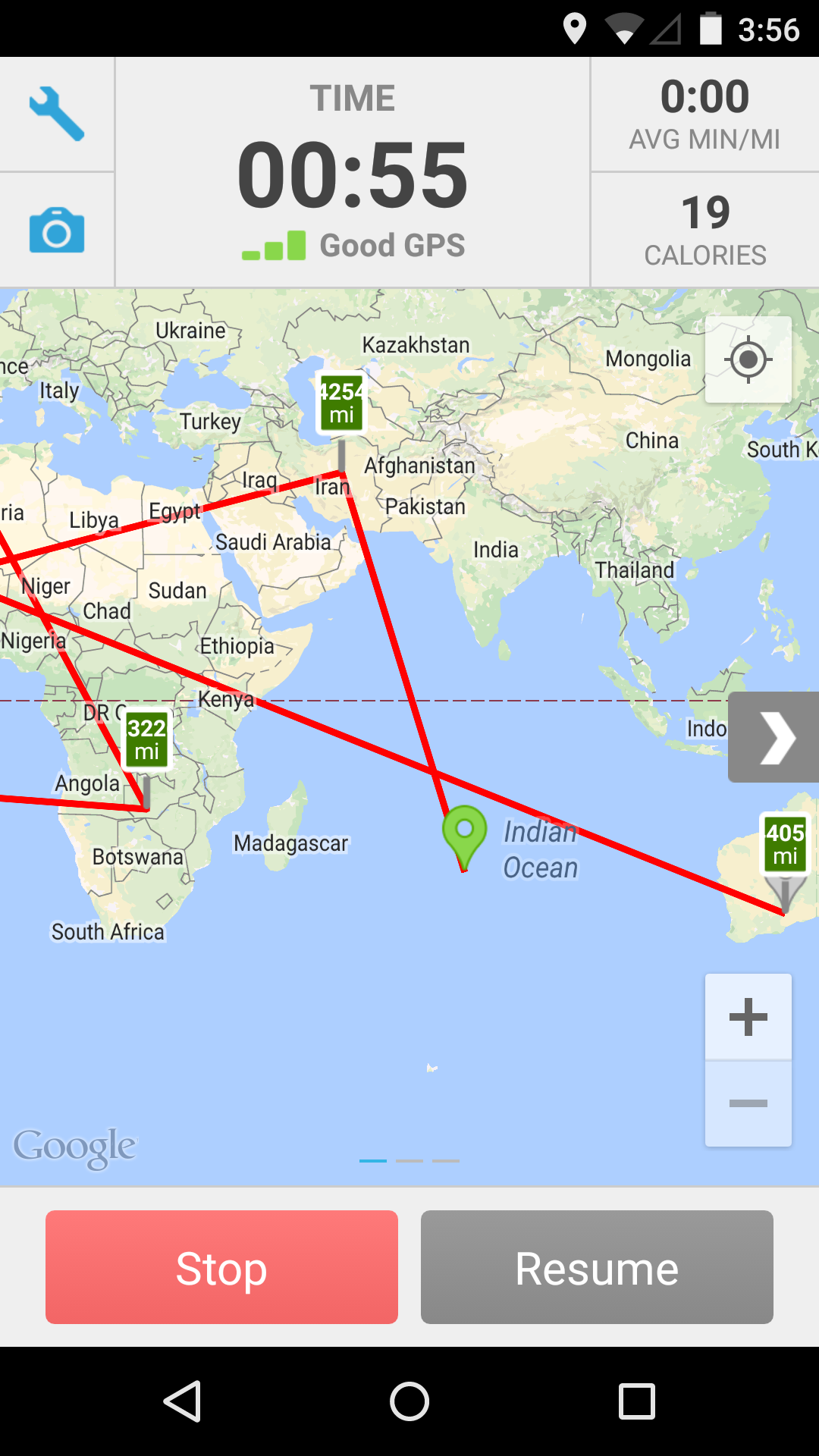}}
  \caption{\textit{RunKeeper} fitness app running in alternate location namespaces.}
  \label{fig:location_app}
\end{figure}

Figure \ref{fig:location_app} shows screenshots of a popular fitness app, \textit{RunKeeper}, that we used to demonstrate the isolated location namespaces. Figure \ref{fig:fuzzy_loc_ns} shows points collected during an activity while the app's \texttt{uid} is assigned the noisy location namespace. Figure \ref{fig:random_loc_ns} shows the same app while assigned to the random location namespace. Note that in both cases, the app's display indicates ``Good GPS'', demonstrating the complete transparency of these namespaces to this unmodified app.

\subsection{Subscriber Information Service} 
\label{subsec:phone}

\texttt{iphonesubinfo} is a hidden service used exclusively by \texttt{TelephonyManager} to service app requests for subscriber information such as IMEI, mobile equipment identifier (MEID), electronic serial number (ESN), phone number, voicemail number, private/public user identities, home network name, etc. Several of these values have significant security and privacy implications and are known to be malware targets \cite{Enck:2011:SAA:2028067.2028088}. Although protected by Android's \texttt{READ\_PHONE\_STATE} permission, misusing or malicious apps can easily legitimize declaration of this permission since it is necessary for a number of common features, such as those provided by \texttt{PhoneStateListener}.

To isolate an untrusted app from or more of the data values returned by \texttt{iphonesubinfo}, we PINPOINTed this system service. We enabled the non-global namespace by modifying the internal telephony \texttt{ProxyController} to instantiate \texttt{PhoneSubInfoController\_1} as well as \texttt{PhoneSubInfoController}. The former starts \texttt{iphonesubinfo\_1} service with an API identical to \texttt{iphonesubinfo}, started by the latter. When an untrusted app is assigned to the alternate \texttt{iphonesubinfo} namespace, it can obtain the same instance of \texttt{TelephonyManager} as trusted apps can, but any subsequent calls to \texttt{getDeviceId()}, \texttt{getLine1Number()}, etc. by the untrusted app are processed by \texttt{iphonesubinfo\_1}. \texttt{iphonesubinfo\_1} returns different values for sensitive subscriber parameters.

When porting this design to Android 5.1, we found that the underlying structure of the telephony service had changed significantly. In particular, the \texttt{ITelephony} (\texttt{phone} service) interface was enhanced to include its own \texttt{getDeviceId()} call, and \texttt{TelephonyManager} was modified to obtain the device ID from this interface rather than \texttt{IPhoneSubInfo} as was the case in 4.4.4. Thus, apps assigned to \texttt{iphonesubinfo\_1} would still get the device's real IMEI because our isolation did not include every object that could return that sensitive data. This necessitates an expansion of the isolation boundary to include both \texttt{iphonesubinfo} and \texttt{phone} services, and is a good example of needing to identify all possible means of access to the sensitive resource related to the end security goal.

\begin{figure}[!t]
  \centering
  \subfigure[\textit{Your SIMCard} running in global \texttt{iphonesubinfo}/\texttt{phone} namespace.]{\label{fig:imei_global}
    \includegraphics[height=1.25in,width=1.25in]{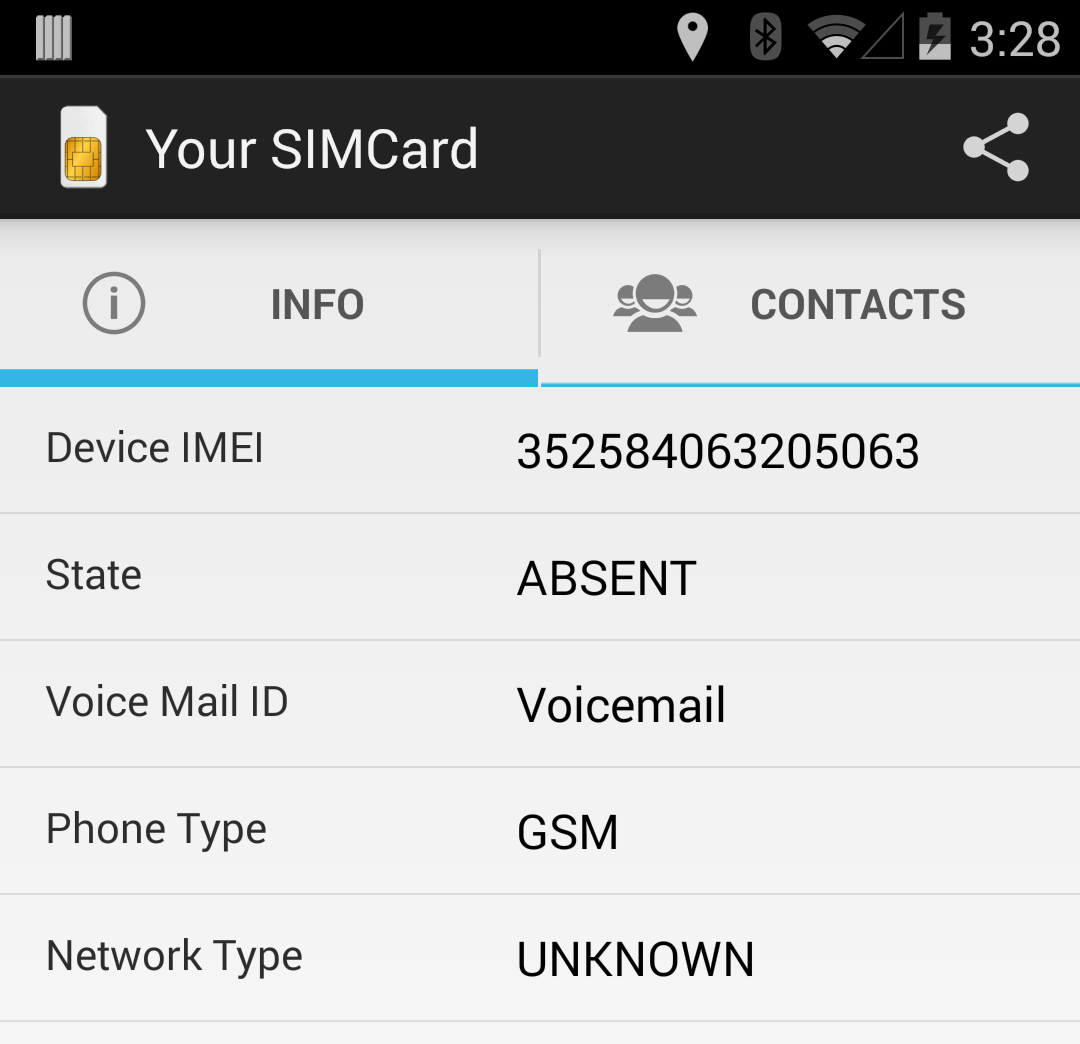}}
  \qquad
  \subfigure[\textit{Your SIMCard} running in alternate \texttt{iphonesubinfo}/\texttt{phone} namespace.]{\label{fig:imei_ns}
    \includegraphics[height=1.25in,width=1.25in]{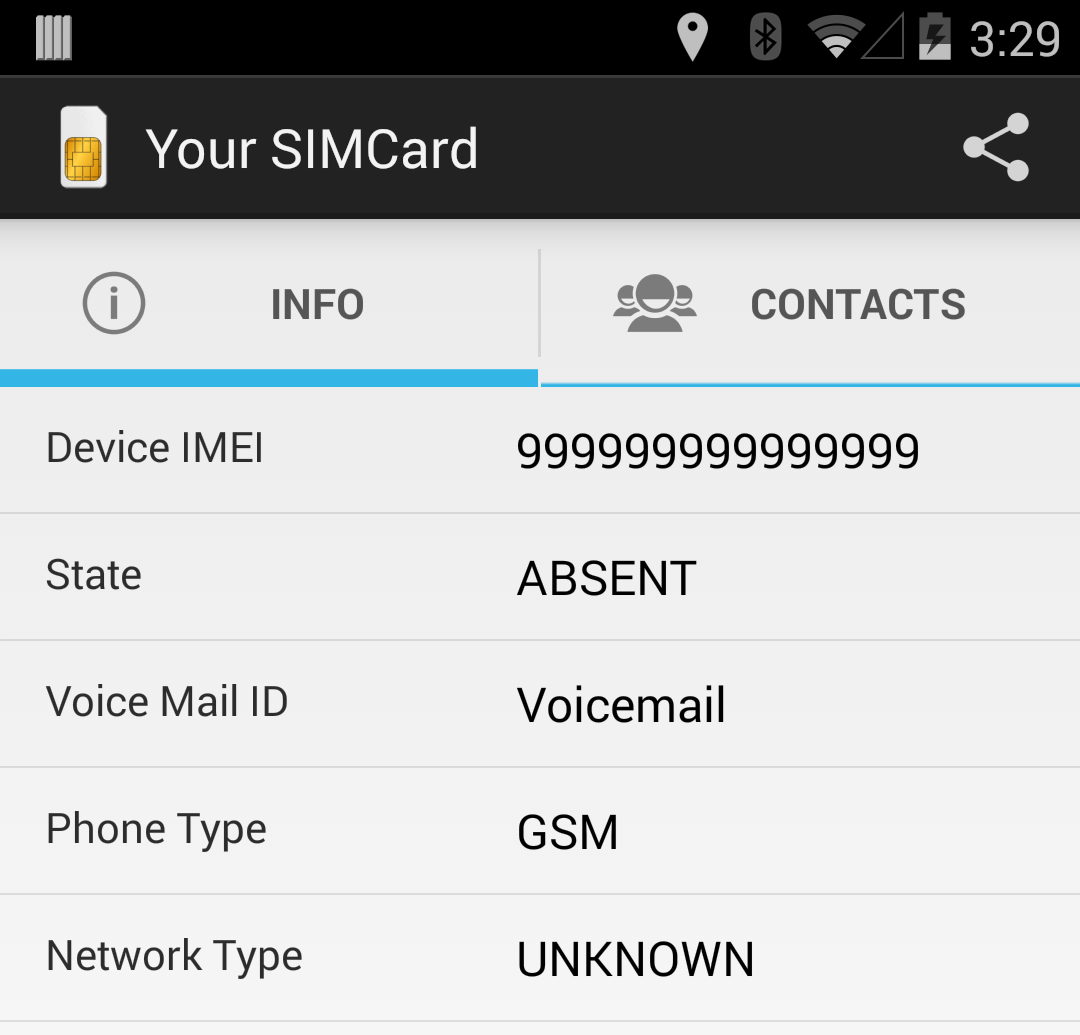}}
  \caption{\textit{Your SIMCard} running in different \texttt{iphonesubinfo}/\texttt{phone} namespaces.}
  \label{fig:imei}
\end{figure}

To demonstrate effectiveness of our PINPOINTed subscriber information service, we obtained the popular app \textit{Your SIMCard}. Figure \ref{fig:imei} shows this app running unmodified in both global (Figure \ref{fig:imei_global}) and fake (Figure \ref{fig:imei_ns}) \texttt{iphonesubinfo}/\texttt{phone} namespaces. In the global namespace, the actual, valid IMEI of our test device is returned, while a fake IMEI is returned to the app after it has been assigned to the alternate \texttt{iphonesubinfo\_1}/\texttt{phone} namespace by adding its \texttt{uid} to the \texttt{nspolicy} file.

\subsection{Input Method Service} 
\label{subsec:ime}

Input Method Editors (IME) are screen controls that enable users to enter text. Currently, there are about 900 third-party keyboard apps published on the Google Play store, with at least 10 having more than one million downloads. Most require \texttt{INTERNET} or \texttt{WRITE\_EXTERNAL\_STORAGE} permissions, which enable the IME to log or transmit any data that's typed in. In an empirical study of keyboard apps, it was found that more than 61\% require three or more permissions giving them the ability to exploit keylogging and man-in-the-middle attack vectors \cite{6680023}. To illustrate this threat, consider sensitive apps like banking or purchasing apps, which often require users to enter bank card numbers or passwords for authentication. All entry of these values is done via the current IME, selected by the user. If the IME is malicious, an attacker can easily collect these values \cite{MMannan:2011:JCS:10.3233}.

\begin{figure}[!t]
\centering 
\includegraphics[width=3in]{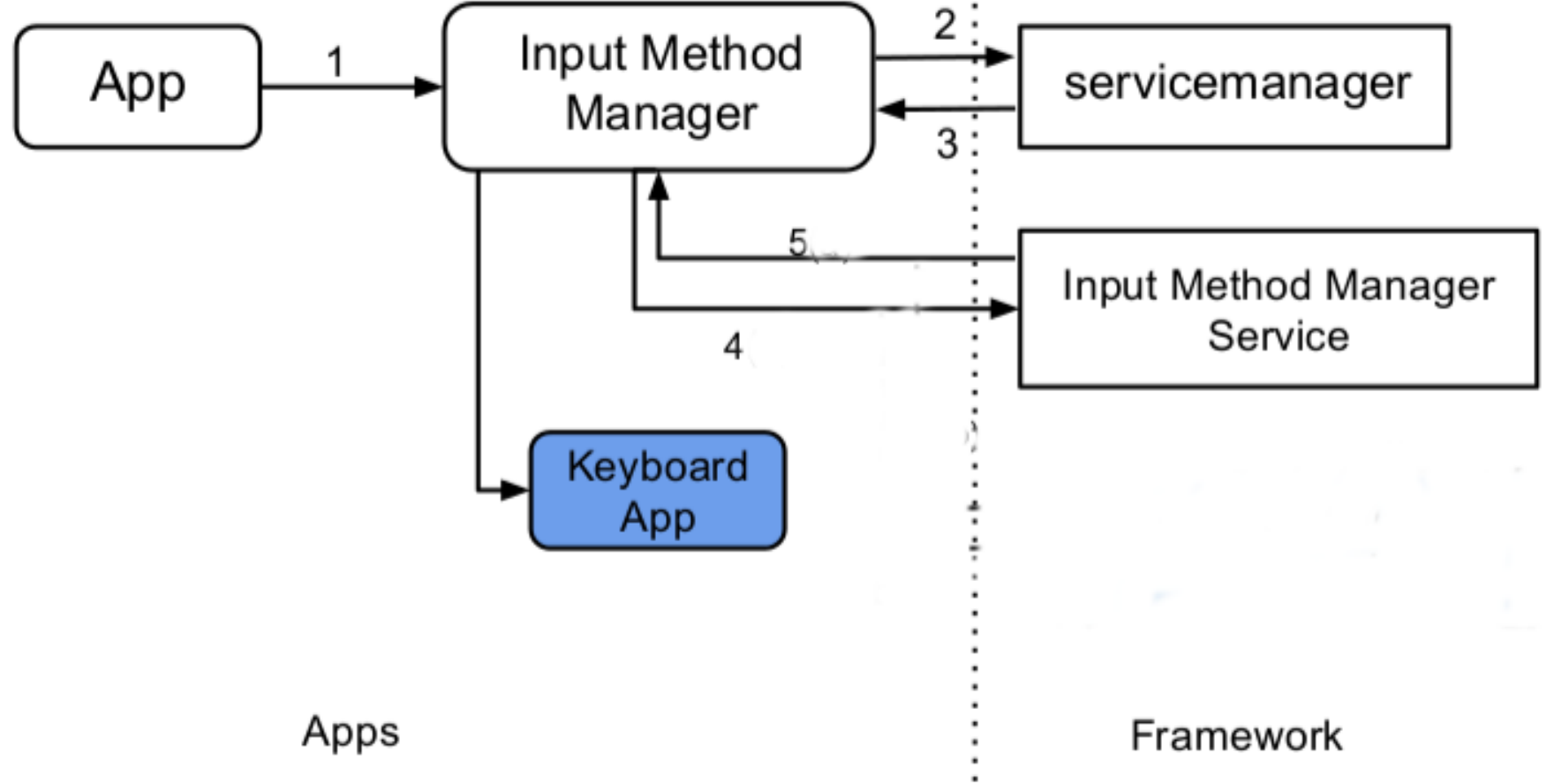}   
\caption{Input method framework architecture.}
  \label{fig:inputmnsarch}
\end{figure}

The overall working architecture of IMEs is shown in Figure \ref{fig:inputmnsarch}. In every application's context space, there exists an instance of \texttt{InputMethodManager} (path 1) which is used to communicate with a system-wide service, \texttt{InputMethodManagerService}. When an input field comes into focus, the app's \texttt{InputMethodManager} invokes this system service (paths 4 and 5) after obtaining its handle via \textit{Service Manager} (paths 2 and 3). With this handle, the app may obtain a unique \texttt{InputConnection} \textit{Binder} token from \texttt{InputMethodManagerService} for making direct calls to the IME keyboard app. Using this token, the system is able to secure and control interactions among multiple applications and multiple IMEs \cite{InputMethodDH}.

Currently, apps do not have control over the IME selected by the user. Instead, the system will bring up the user's selected IME whenever any text field comes into focus. While Google has recognized the security and privacy issues associated with this design \cite{InputMethodManager}, the current measures rely on the user to make wise choices regarding IME installation and selection. Using session information attached to each window instance by \texttt{WindowManager}, the Input Method Framework (IMF) ensures that only the active activity can get access to the data being entered. Furthermore, \texttt{InputMethodManagerService} ensures that all messages received from running IME applications are from the current user. Importantly, this includes messages for changing IMEs (i.e., messages resulting from calls to \texttt{InputMethodManager.setInputMethod()}), which are guarded with the token to ensure that they originated from explicit user selection. However, none of these protections will help if the IME itself is malicious or compromised and the user selects it.

\begin{figure}[!t]
\centering
\includegraphics[width=3in]{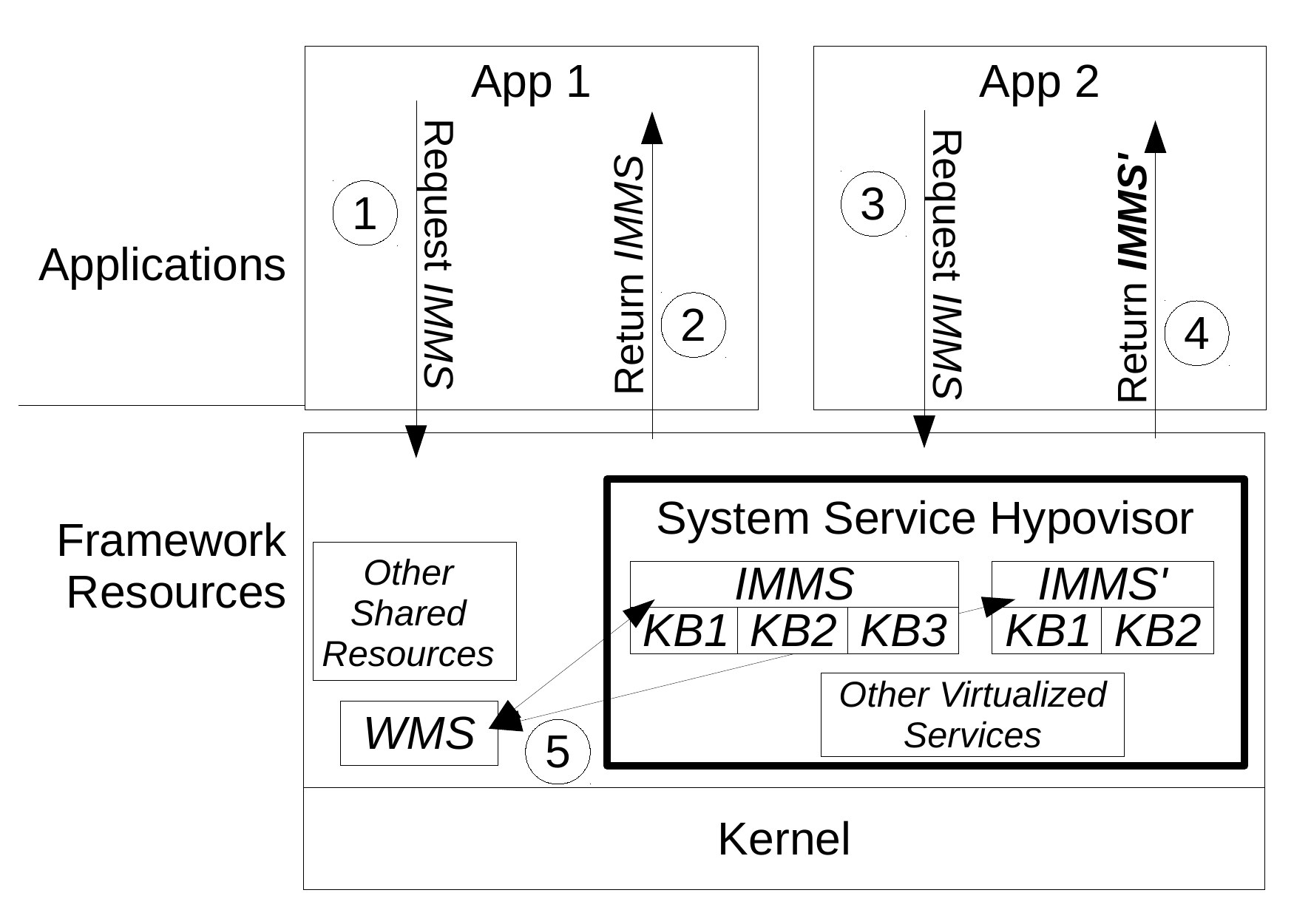}   
\caption{PINPOINTing \texttt{InputMethodManagerService}.}
  \label{fig:inputmns}
\end{figure}

By PINPOINTing the \texttt{InputMethodManagerService}, we are able to provide a mechanism to shield sensitive apps from falling victim to a malicious IME selected by a tricked user. Figure \ref{fig:inputmns} shows the PINPOINT concept applied to input methods. This is accomplished by using the our PINPOINT service hypovisor prototype to virtualize \texttt{InputMethodManagerService}. In the figure, $IMMS$ corresponds to the ``real'' \texttt{InputMethodManagerService} (\texttt{input\_service}), while $IMMS'$ is a second service (\texttt{input\_service\_1}), with an identical interface and features except for the fact that it holds only a subset of all available IMEs.

As suggested in Section \ref{sec:concept}, there are additional complexities with virtualizing IMEs due to dependencies with other objects in the system. Because of interactions with \texttt{WindowManager} mentioned above, we needed to make minor modifications to \texttt{WindowManager}, so that it can be aware of the all the \texttt{InputMethodManagerService} namespaces running and push updates about the current activity to all of them. As with location service, this situation corresponds to dependent resource $G$ in Figure \ref{fig:concept}. To enable independent \texttt{InputConnection} from each app's \texttt{InputMethodManager} instance to each service, we created a Java interface which all of the \texttt{InputMethodManagerService} instances implemented.

\begin{figure}[!t]
  \centering
  \subfigure[Non-critical app running in global IME namespace, showing all input methods, including a 3\textsuperscript{rd} party (\ding{172}), as selection options.]{\label{fig:inputeval-1}\includegraphics[width=1.5in]{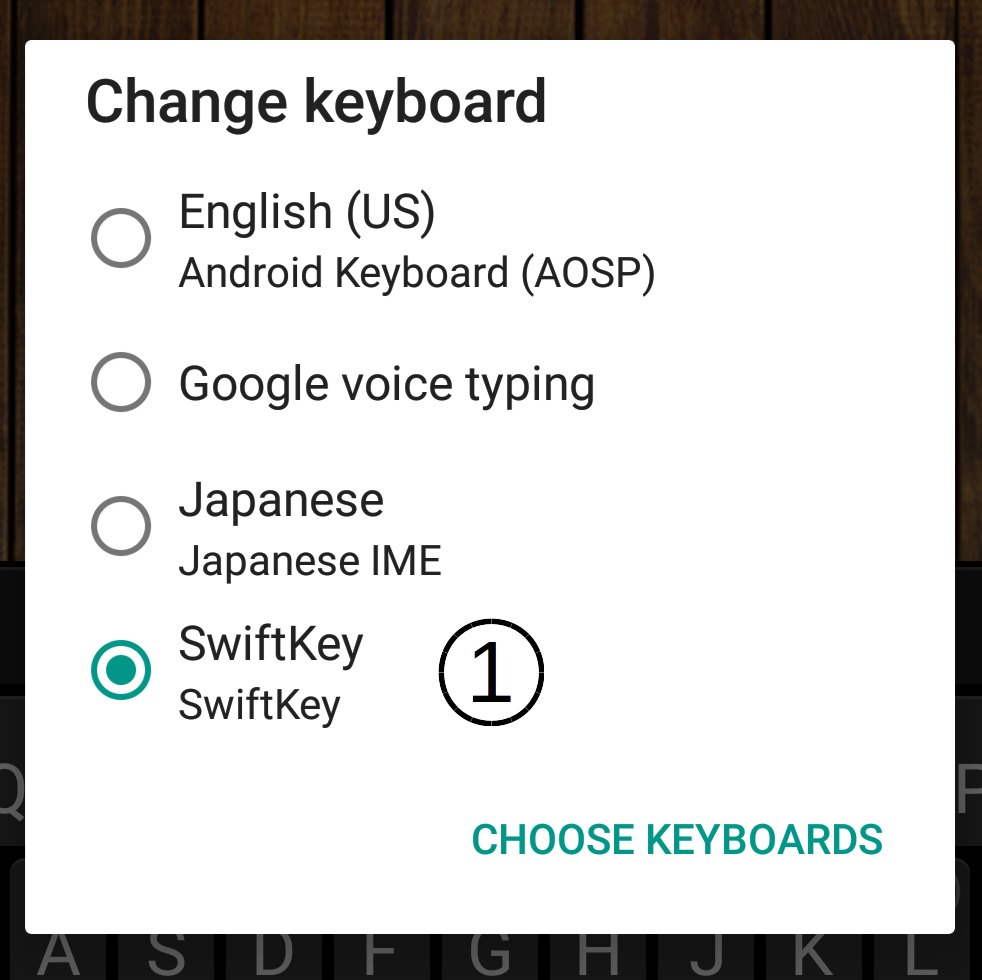}}
  \qquad
  \subfigure[Critical banking app running in alternate IME namespace, showing only built-in input methods as selection options.]{\label{fig:inputeval-2}\includegraphics[width=1.5in]{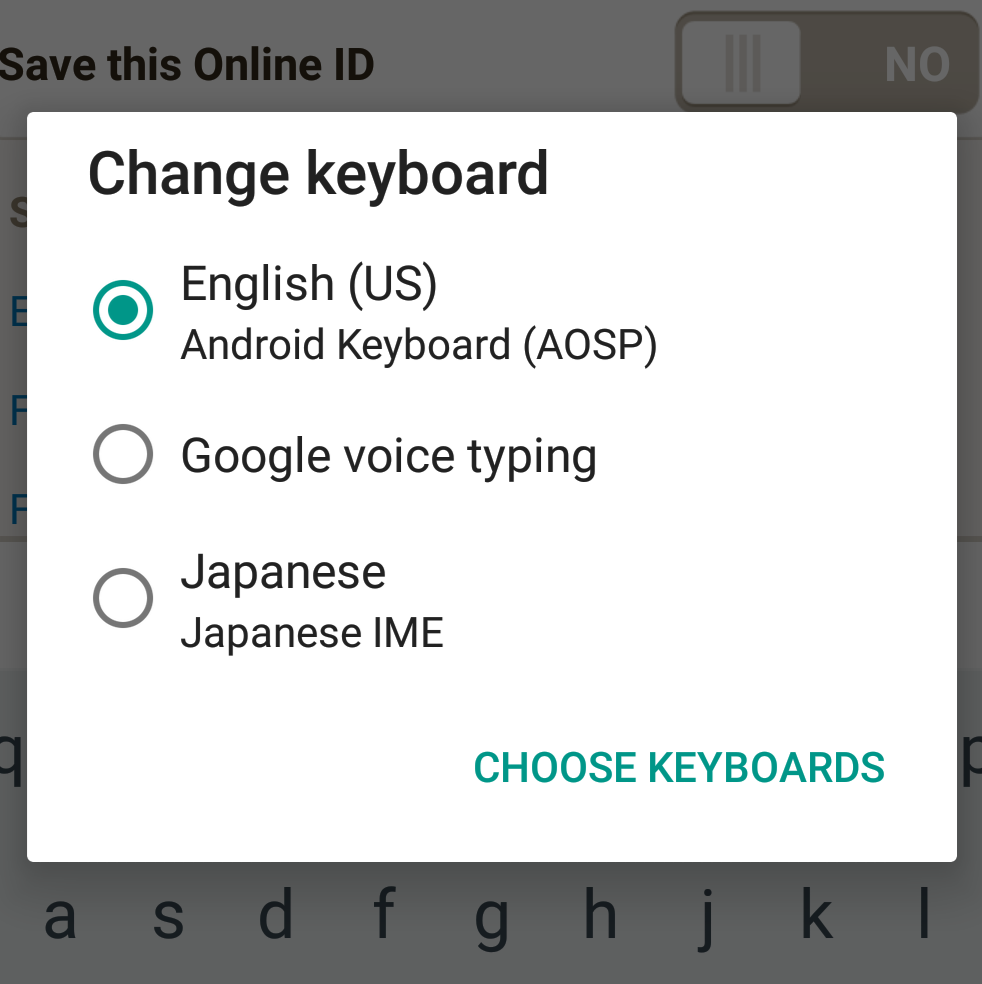}}
  \caption{Non-critical and critical apps running in different IME namespaces.}
  \label{fig:inputeval}
\end{figure}

A demonstration of our IME namespaces is shown in Figure \ref{fig:inputeval}. Figure \ref{fig:inputeval-1} depicts a non-critical app, \textit{EatStreet}, assigned to the global IME namespace, where any IME can be used, including a representative untrusted IME, \textit{SwiftKey} (added from Google Play). Here, the \textit{Choose input method} dialog shows all installed input methods. In contrast, the critical banking app of Figure \ref{fig:inputeval-2} has been assigned to the alternate IME namespace in order to protect its data from possible malicious IMEs. Hence, the chooser only allows selection of trusted IMEs, while \textit{SwiftKey} is excluded as an authorized IME for this app.

\subsection{Sensor Service} 
\label{subsec:sensor}

Modern mobile devices have a rich set of environmental and motion sensors available to apps. Unfortunately, the Android security architecture does not extend to most of these sensors, making it all too easy for malware to utilize them to compromise user data entry \cite{Xu:2012:TIU:2185448.2185465}\cite{Aviv:2012:PAS:2420950.2420957}, eavesdrop on voice communications \cite{6680832}, track user movements, and infer location \cite{Komeda:2014:UAR:2638728.2641299}. By PINPOINTing \texttt{SensorService}, we enable the user to take advantage of apps without needing to also trust their handling of sensor data.

In the Android platform, apps may acquire sensor data by getting an instance of \texttt{SensorManager}, which in turn accesses raw sensor data via \texttt{SensorService}, a native system service. \texttt{SensorService}'s \texttt{threadLoop()} collects raw sensor data in a structured data buffer of type \texttt{sensor\_event\_t}, which is then returned to the app via its \texttt{SensorManager}'s \texttt{SensorEventConnection}. The buffer structure contains raw sensor data for each of the device's sensors including acceleration, magnetic, orientation, gyro, temperature, distance, light, pressure, and relative humidity.

To PINPOINT sensor resources, we followed the same general approach as with previous examples, by adding two additional native \texttt{SensorService}s to the device, and registering them with \textit{Context Manager} as \texttt{sensorservice\_1} and \texttt{sensorservice\_2}. For demonstration purposes, we hardcoded \texttt{sensorservice\_1} to overwrite the gyro, magnetic, and orientation structure members of the buffer structure with random data before it is returned to the app's \texttt{SensorManager}. Likewise, \texttt{sensorservice\_2} is hardcoded to overwrite only the light structure member of the structure with random values. Structure members containing data from other sensors are passed through unmodified.

\begin{figure}[!t]
\centering
\includegraphics[height=2.5in]{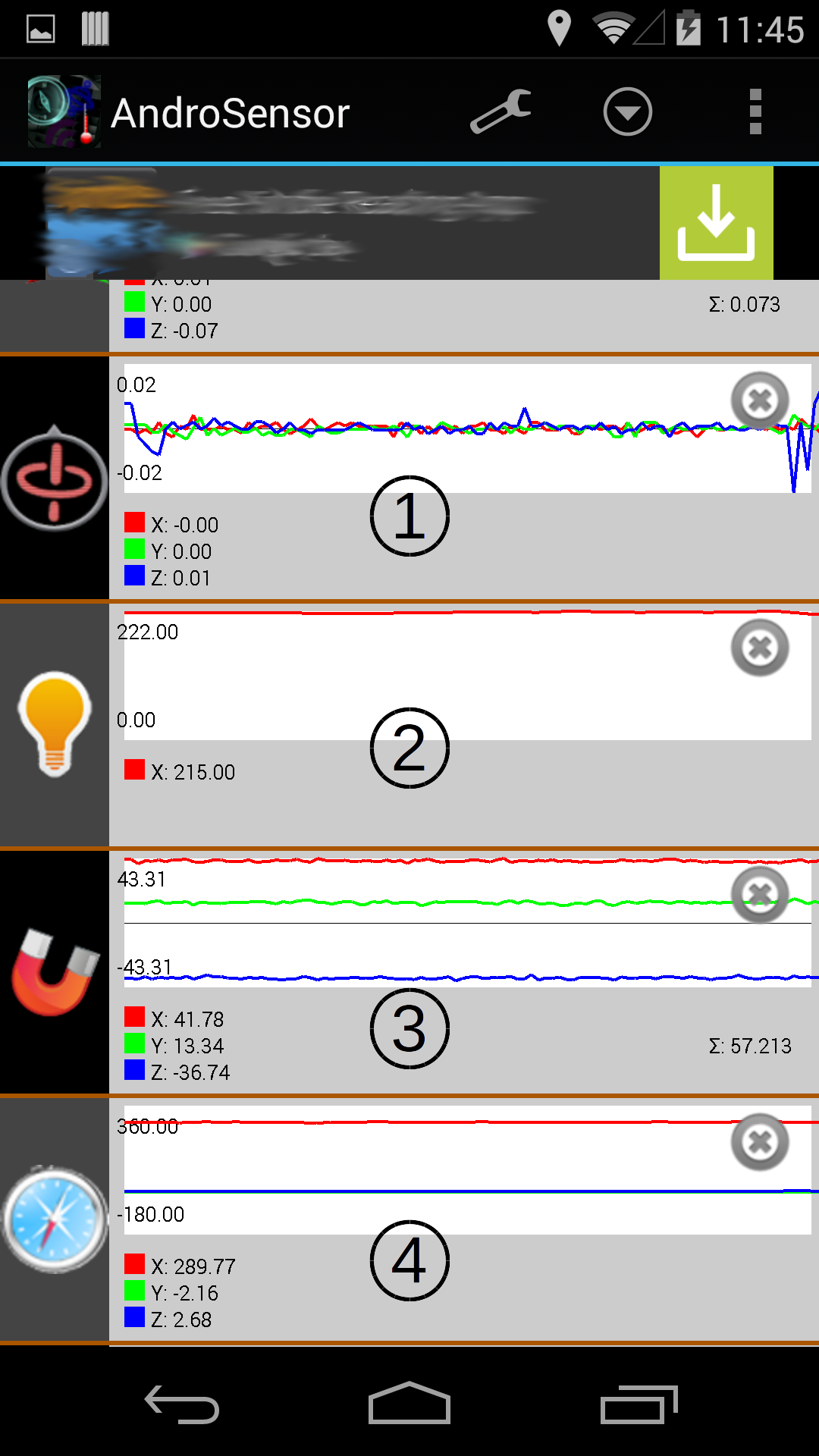}   
\caption{\textit{AndroSensor} running in global sensor namespace showing normal traces for gyro (\ding{172}), light (\ding{173}), magnetic (\ding{174}) and orientation (\ding{175}) sensors.}
\label{fig:sensorGlobal}
\end{figure}

\begin{figure}[!t]
  \centering
  \subfigure[\textit{AndroSensor} running in 1\textsuperscript{st} alternative sensor namespace showing normal trace for light  sensor (\ding{173}), and random traces for gyro (\ding{172}), magnetic (\ding{174}) and orientation (\ding{175}) sensors.]{\label{fig:sensor_1}\includegraphics[height=2.5in]{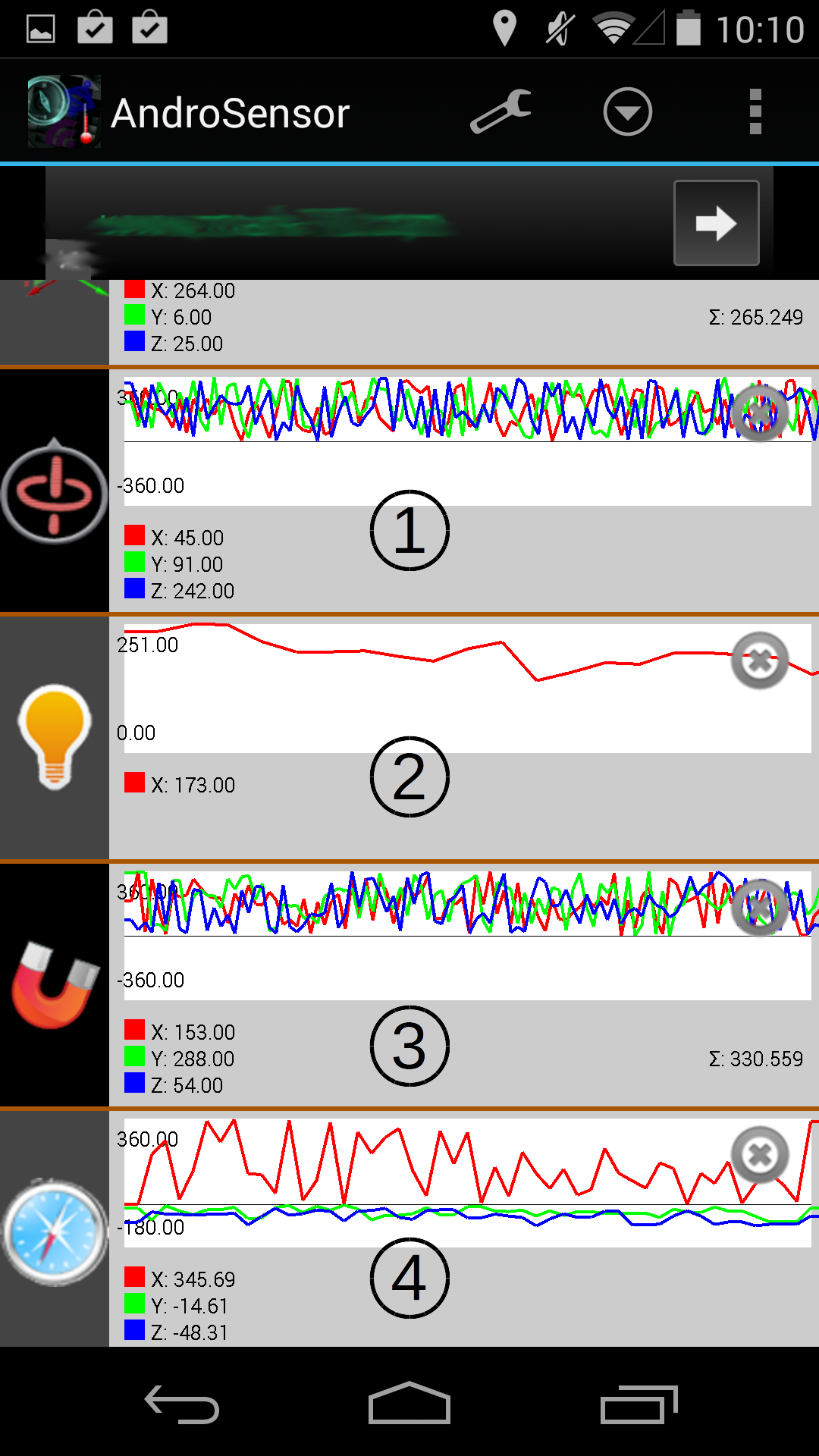}}
  \qquad
  \subfigure[\textit{AndroSensor} running in 2\textsuperscript{nd} alternative sensor namespace showing normal traces for gyro (\ding{172}), magnetic (\ding{174}), and orientation (\ding{175}) sensors, and random trace for light sensor (\ding{173}).]{\label{fig:sensor_2}\includegraphics[height=2.5in]{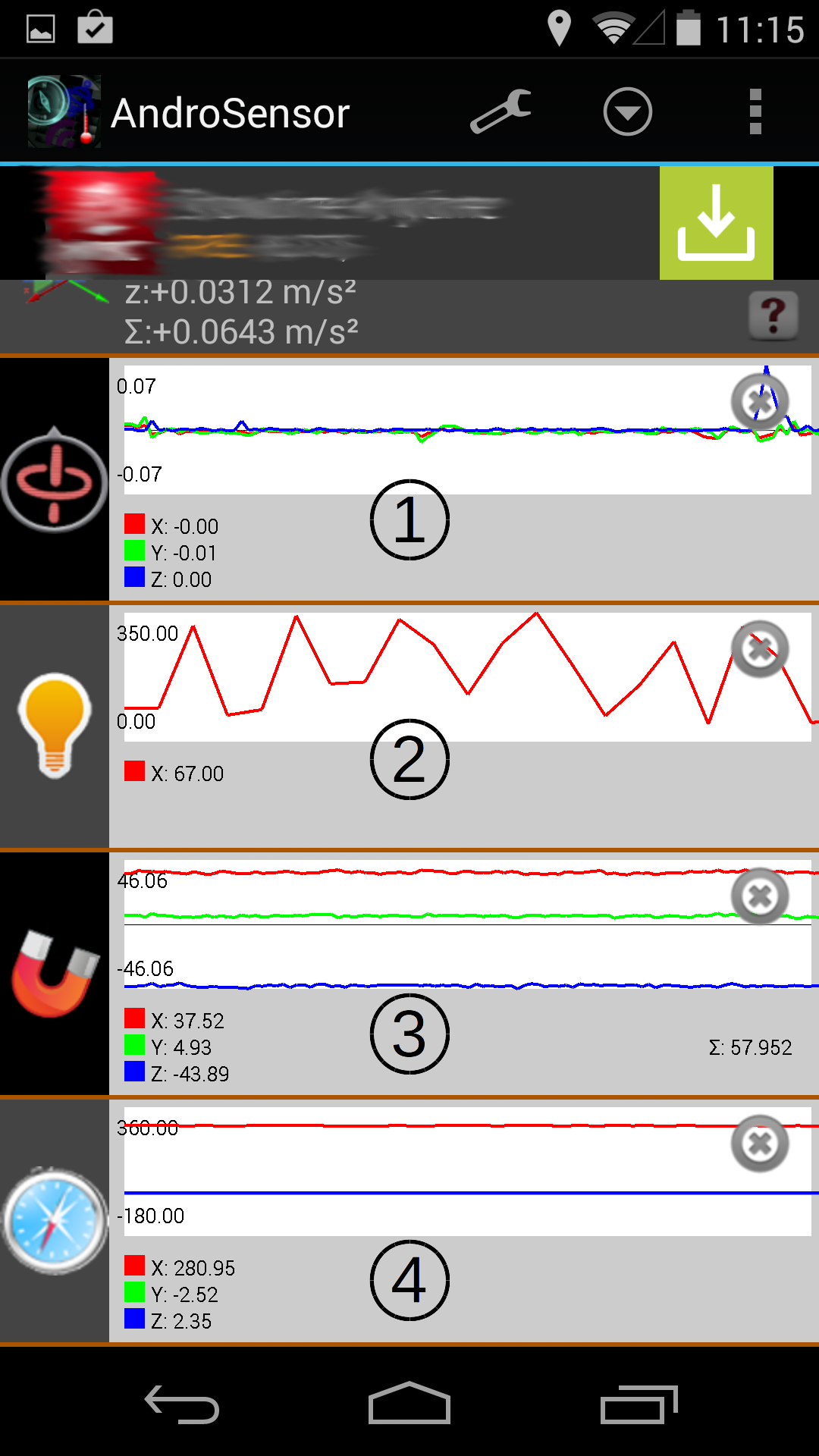}}
  \caption{\textit{AndroSensor} running in alternative sensor namespaces.}
  \label{fig:sensor}
\end{figure}

With three possible sensor service handles on the device, \texttt{SensorManager}s of apps assigned to one of the two alternate sensor namespaces are always given handles to \texttt{sensorservice\_1} or \texttt{sensorservice\_2}, depending on their assignment. To demonstrate the effectiveness of this, we downloaded \textit{AndroSensor} a popular Google Play Store app, and ran it in each of the three sensor namespaces. Figure \ref{fig:sensorGlobal} shows \textit{AndroSensor} running in the global sensor namespace, with all sensor traces steady, indicating a stable physical environment. In contrast, Figures \ref{fig:sensor_1} and \ref{fig:sensor_2} show \textit{AndroSensor} running in the alternate sensor namespaces of \texttt{sensorservice\_1} and \texttt{sensorservice\_2}, respectively. For all three cases, the physical envronment was approimately the same.

\section{Evaluation}
\label{sec:evaluation}

\subsection{Performance}
\label{subsec:perf}

\begin{figure}[!t]
  \centering
  \subfigure[Average LINPACK CPU performance score vs. number of namespaces.]{\label{fig:cpumeasure}
    \includegraphics[width=0.2\textwidth]{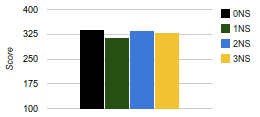}}
    \qquad
  \subfigure[Average file I/O performance score vs. number of namespaces (Quadrant file I/O).]{\label{fig:iomeasure}
    \includegraphics[width=0.2\textwidth]{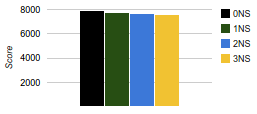}}
  \qquad
  \subfigure[Average graphics performance score vs. number of namespaces (Quadrant 2D \& 3D).]{\label{fig:graphicsmeasure}
    \includegraphics[width=0.2\textwidth]{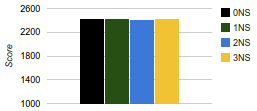}}
    \qquad
  \subfigure[Average browser performance score vs. number of namespaces (SunSpider).]{\label{fig:jsmeasure}
    \includegraphics[width=0.2\textwidth]{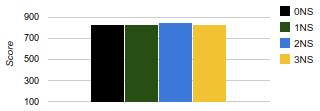}}
  \caption{Benchmarking results for 0-, 1-, 2- and 3-namespace configurations.}
  \label{fig:perfEval}
\end{figure}

To evaluate the overall performance impact of PINPOINTing services, we performed the benchmark tests shown in Table \ref{table:benchmarks}, with and without namespaces. For each benchmark, we measured performance under four different device configurations: \textit{0NS} represents stock Android without any PINPOINT capability or namespaces, while \textit{1NS}, \textit{2NS}, and \textit{3NS} represent devices configured with one, two and three PINPOINTed services, respectively. Figure \ref{fig:perfEval} shows the average value of 10 runs of each benchmarking test.

\begin{table}[!t]
\centering
\renewcommand{\arraystretch}{1.3}
\caption{Evaluation benchmarks used.}
\label{table:benchmarks}
\centering
\begin{tabular}{c||c||c}
\hline
\bfseries Name & \bfseries Version & \bfseries Workload type\\
\hline\hline
Linpack & 1.2.8 & CPU\\
\hline
Quandrant Advanced Edition & 2.1.1 & File I/O \\
\hline
Quandrant Advanced Edition & 2.1.1 & 2D \& 3D \\
\hline
SunSpider & 1.0.2 &  CPU \& I/O\\
\end{tabular}
\end{table}

\begin{figure}[!t]
  \centering
  \includegraphics[width=2.5in]{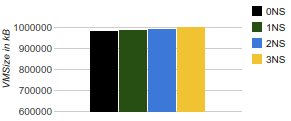}
  \caption{Average memory footprint in kB (\texttt{VmSize}) for 0-, 1-, 2- and 3-namespace configurations.}
  \label{fig:memEval}
\end{figure}

We also measured the impact on memory of adding PINPOINTed services. Since each running service represents additional threads within \texttt{SystemServer}, we measured \texttt{VmSize} of the \texttt{system\_server} process by reading its \path{/proc/<pid>/status} under each of the same four configurations. Figure \ref{fig:memEval} shows the average value of 10 measurements of memory footprint for each configuration.

\subsection{Discussion}
\label{subsec:discussion}

Performance evaluation results presented in Section \ref{sec:evaluation} indicate that increasing numbers of PINPOINTed services has no apparent effect on CPU, browser, or graphics performance. On the other hand, we observe a clear correlation between the number of PINPOINTed services and file I/O. Decreases in this score with increasing numbers of namespaces is expected due to an increase in policy file size and associated data structures being parsed and searched by \texttt{servicemanager} during every service lookup request in order to support namespace reassignments of running apps. In our current, unoptimized design, file I/O performance degrades by an average of 1.57\% of the \textit{0NS} value for each additional namespace represented in the policy file. Although this degradation is negligible for simple policy files, we feel that this is an area for improvement. Our future implementations will include an optimization of this code, and policy options to configure how often policy lookups are performed.

We also observed a growth in \texttt{system\_server}'s memory size that is correlated to the the number of additional service objects (i.e., namespaces) available for use in the \texttt{system\_server} process. On average, we observed this increase to be approximately 0.64\% of the \textit{0NS} value per each additional service. For a system with one additional IMEI namespace, two additional location namespaces, one additional input method namespace, and two additional sensor namespaces, \texttt{system\_server} would have an approximately 3.84\% larger memory footprint than the stock process. Note that an unused namespace still consumes additional memory, but since it does not add to the policy file, it will not contribute to file I/O degradation.

\section{Related Work}
\label{sec:related_work}

A number of previous efforts have addressed the problem of untrusted apps having access to sensitive or private information. Some of these address specific types of data, such as location, while others look for more general solutions.

Two significant isolation approaches that influenced us tremendously are Cells\cite{Dall:2012:DIE:2324876.2324877} and AirBag \cite{wu2014airbag}. Cells leverages Linux Namespaces to allow multiple Android user spaces to run simultaneously on a single hardware platform. Each user space, or virtual phone (VP), is isolated in a combination of separate Linux Namespaces for file system paths, process identifiers, IPC identifiers, network interface names, user names, and hardware devices. Cells introduces the concept of a foreground and multiple background phones that are isolated from each other so that malicious or buggy apps in one VP cannot affect others. Isolation in Cells is thus achieved at the virtual phone boundary.

AirBag also leverages Linux Namespaces, but achieves isolation at the native runtime boundary. This is accomplished by instantiating a separate app runtime that has virtually no interaction with the original native runtime. Each isolated runtime contains its own copies of key service processes and daemons, such as \texttt{vold}, \texttt{binder} and \texttt{servicemanager} that are launched in separate namespaces as compared with the normal runtime. Thus, an untrusted app ``sees'' an entirely different set of services, binder objects, file paths, etc. through the lens of its decoupled runtime. The untrusted app cannot communicate with apps in different runtimes, and the system resources it can view and control are completely dictated by the isolated runtime.  Condroid\cite{condroid2015} improves on AirBag's design by restoring binder communications via virtual binders and increasing efficiency by enabling many system services to be shared among runtimes instead of duplicated.

While Cells, Airbag and Condroid provide excellent general-purpose isolation, their designs are complex, burden the system, and somewhat intrusive from the user's point of view. For example, all three require special modifications to numerous shared hardware drivers, duplication of system processes and resources not related to the security goals, and introduce significant usability restrictions. Although they leverage lightweight Linux Namespace isolation, key benefits of Namespaces (Table \ref{table:traits}) are lost when the Android Framework is added on top since many fundamental aspects of Android's open design are broken by the kernel-level isolation. Fixing these problems greatly complicates the designs. Thus, our main difference from these works is our deliberate choice \textit{not} to provide a general-purpose solution, but rather one that addresses specific security goals by directly isolating the specific Framework objects associated with the security goals. For these specific cases, our approach is simpler, less burdensome, and more usable.

MOSES \cite{Russello:2012:MSO:2295136.2295140} is a framework designed to isolate applications and data for the purpose of protecting sensitive corporate data. While MOSES also represents an effective general solution to securing corporate data leaks on mixed-use personal/business devices, it is not very suitable for protecting users' privacy or securing specific resources because of its security profile-centric architecture that forces explicit switching and carries performance penalties. 

IPC Inspection \cite{Felt:2011:PRA:2028067.2028089} and Quire \cite{Dietz:2011:QLP:2028067.2028090} prevent privilege escalation and confused deputy attacks among apps and do not address an app's direct access to resources it already has adequate permissions for, as our work does. TaintDroid \cite{Enck:2010:TIT:1924943.1924971} inspects and analyzes information flows across the system, but does not provide the means to manipulate or block information. AppFence \cite{Hornyack:2011:TAD:2046707.2046780} leverages TaintDroid monitoring to enable data substitution and blocking. For information resources, such as location and IMEI, the resulting capability is similar to some of our basic namespaces. However, AppFence cannot control the semantics of functional resources, such as we demonstrated with the input method namespace. Furthermore, AppFence's substitution and blocking capabilities affect information resources for the entire platform rather than being selective for individual apps as is the case in our system services case study.

Mr. Hide \cite{Jeon:2012:DAM:2381934.2381938} adds finer-grained permissions to apps by way of byte code rewriting, while APEX \cite{Nauman:2010:AEA:1755688.1755732} introduces context-sensitive run-time permissions. Compac \cite{Wang:2014:CEC:2557547.2557560} allows different components within apps, such as in-app ads, to have different sets of permissions. These and other permission-enhancement designs can only restrict access to resources and are unable to redefine them as we do. 

Finally, as location data is widely viewed as having serious privacy implications, there are numerous works specific to improving location privacy. LP-Guardian \cite{fawaz2014location}, LISA \cite{6682719}, and Koi \cite{Guha:2012:KLP:2228298.2228317} are examples of these. While each is effective for controlling or preventing the use of location data, they are not generally applicable to other resources as PINPOINT is. As our case study on PINPOINTing system services demonstrates, if the point of virtualization is chosen wisely, the resulting isolation capability is flexible enough to apply to classes of resources rather than only specific ones as these works do.

\section{Future Directions}
\label{sec:future}

In our present work, we have gained a tremendous insight into the trade-off between isolation design alternatives, system complexity, usability, convenience and effectiveness. We plan to further quantify these relationships so that we can make informed choices when addressing the high-level requirements typically stated by end-users. Ultimately, we plan to formalize the PINPOINT methodology, so that security designers can easily understand the trade space of PINPOINT designs vs. general-purpose approaches.

Through our case study of implementing a services hypovisor, we've acquired a sense for the difficulty of implementing a representative PINPOINT hypovisor and its companion virtual resources within the Android Framework. Encouraged by our experiences, we plan to consider the potential benefits of PINPOINTing other resources, including those outside the purview of \textit{Service Manager}. Following from this, we envision implementing a container abstraction, whereby multiple, heterogeneous PINPOINTs, Linux Namespaces, and other forms of access control and virtualization can be easily combined by the end-user to form easily-understood security and privacy macros such as ``incognito,'' ``banking,'' etc.

We recognize the inflexibility of having to define PINPOINTed resources at system build time. As such, we see opportunities to investigate techniques for establishing new PINPOINTs while the system is running. Also, we would like to evolve our current rudimentary means of policy configuration into a more powerful and intuitive means for end-users to configure, combine and use PINPOINTed resources, possibly through an advanced launcher interface.

Additional project details, status, and instructions for requesting access to our prototype code are available at \path{https://goo.gl/2pJeMp}.

\section{Conclusion}
\label{sec:conclusion}
We have presented PINPOINT, a Android resource isolation strategy that forgoes general-purpose solutions in favor of a ``building block'' approach, similar in concept to Linux Namespaces, but implemented in the Android Framework. By addressing stated security goals and no more, PINPOINT yields an effective result using only the minimum amount of isolation. This helps minimize or eliminate the negative side-effects that are sure to emerge when large parts of Android's open architecture are subject to isolation.

Through our case study on Android System Services, we uncover the primary considerations of a PINPOINT designer. These include correlating the stated security goals with specific Android resources, identifying all Framework objects that have access to these resources, and finding any dependent resources that may exist in the remainder of the system. With this insight, an appropriate hypovisor is then implemented, corresponding objects with alternate semantics are created, and system MACs are updated as necessary to enforce the hypovisor's authority.

Our system services prototype and experiments with location, subscriber, input method and sensor services demonstrate that for specific security or privacy goals, PINPOINT yields effective solutions for unmodified apps with a minimal amount of design complexity, system modifications, or negative impacts on user experience.

\section*{Acknowledgement}
\label{sec:acknowledgement}
We thank the anonymous reviewers for their careful reading of our manuscript and the many insightful comments and suggestions. This work was supported in part by NSF Grant 1318814, and AFRL project GAIHCYBR. Any opinions, findings, conclusions or recommendations expressed in this material are solely those of the authors and do not necessarily reflect the views of the NSF or the US Air Force.

\bibliographystyle{IEEEtranS}
\bibliography{IEEEabrv,refs}


\end{document}